\documentclass[a4paper,twocolumn,superscriptaddress]{revtex4-1}
\usepackage{amsmath}
\usepackage{float}
\usepackage{tabularx}
\usepackage{graphicx}
\newcommand{\nn}{\nonumber\\}

\begin{document}

\title{Accurate and efficient description of interacting carriers in 
quantum nanostructures by selected configuration interaction and
perturbation theory}
\author{Moritz Cygorek}
\affiliation{Department of Physics, University of Ottawa, Ottawa, Canada}
\author{Matthew Otten}
\affiliation{Center for Nanoscale Materials, Argonne National Laboratory, 
Lemont, Illinois, USA}
\author{Marek Korkusinski}
\affiliation{National Research Council of Canada, Ottawa, Canada}
\author{Pawel Hawrylak}
\affiliation{Department of Physics, University of Ottawa, Ottawa, Canada}

\begin{abstract}
We present a method 
to calculate many-body states of interacting carriers in 
million atom quantum nano\-structures based on atomistic tight-binding 
calculations and a combination of iterative selection of configurations and 
perturbation theory.
This method enables investigations of large excitonic complexes and 
multi-electron systems with near full configuration interaction 
accuracy, even though only a small subspace of the full many-body Hilbert space 
is sampled, thus saving orders of magnitudes in computational resources.
Important advantages of this method 
are that the convergence is controlled by a single 
parameter, the threshold, and that ground and excited states can be treated on 
an equal footing. 
We demonstrate the extreme efficiency of the method
by numerical studies of complexes composed of up to 13 excitons, which
requires filling of states up to the fourth electronic shell.
We find that the method generally converges fast as a function of the
threshold, profiting from a significant enhancement 
due to the perturbative corrections.
The role of the choice of single-particle basis states is discussed. It is 
found that the algorithm converges faster in the Hartree-Fock basis only for
highly charged systems, where Coulomb repulsion dominates.
Finally, based on the observation that
second order perturbative energy corrections 
only depend on off-diagonal elements of the many-body Hamiltonian, 
we present a way to accurately calculate many-body states that 
requires only a relatively small number of Coulomb matrix elements.
\end{abstract}

\maketitle

\section{Introduction}
Quantum nanostructures like quantum dots\cite{book_QD,book_SQD, Arakawa_QD, 
QD_RevModPhys}, quantum rings\cite{CdSe_rings}, and 
nanoplatelets\cite{Bayer_nanoplatelets}
are workhorse systems for the development of 
semiconductor-based quantum technology devices, such as
single-photon emitters\cite{singlephoton_Michler,singlephoton_Cosacchi,
singlephoton_Santori} 
or sources of entangled photon pairs\cite{entangled_Orieux,
entangled_Stevenson2006,Dalacu_entangled_2014,
Dalacu_entangled_2019,concurrence_Cygorek, concurrence_Tim_PRL, 
concurrence_Tim, Korkusinski_photon_cascades}.
Due to the confinement of electrons to a small volume, quantum dots
can be viewed as artificial atoms. When two or more atoms are brought 
together they form molecules. Similarly, more complex devices can
be built from quantum dots by fabricating systems with multiple dots that
are close enough to introduce inter-dot tunneling\cite{Bayer_Science_molecules,
QDM_Zielinski, Zielinski_molecules}.
These systems can be used, e.g., to realize 
two- or three-dot spin qubits\cite{spin-spin_nphys,S-T-qubits_science,
Burkard_doubledot2018, Burkard_threespin}.
Complexity is also added when
a quantum dot is loaded with multiple charge carriers\cite{HighlyChargedQD,
QD_charging_exp, ChargedQring_Warburton,ChargedQD_optical_Gershoni}.
In analogy to transition metal elements, occupation of dots with 
multiple electrons can lead to the formation of correlated magnetic states for
partially filled shells\cite{Wojs_multichargedGS}. 
In quantum dots, multi-excitonic complexes\cite{Korkusinski_2X3X}
are interesting, as they can be easily probed by photoluminescence at high 
intensities\cite{HiddenSymmetry_Exp,
Raymond_state_filling_1996, Molinari_few-particle_QDs}.
The biexciton is particularly relevant for the generation of entangled
photon pairs in the biexciton-exciton cascade\cite{concurrence_Cygorek} 
and the emission from the 
lowest-energetic three-exciton complex, which necessarily involves occupation
of the p-shell, contains information about the lateral confinement in the
quantum dot\cite{Zielinski_selfassembled_InAs,MultiXspectra_exp}. 
Similarly, the d- and f-shells can be probed via
the emission of the lowest-energetic states
of seven- and thirteen-exicton complexes, respectively.
Charged excitonic complexes like trions can be used for the generation 
of highly entangled photon cluster states \cite{clusterstate_proposal,
clusterstate_measured} that are required for measurement-based quantum 
computation\cite{Oneway_qcomputer}.
The proposal of implementing a synthetic Haldane 
chain\cite{synthetic_haldane_chain}, which possesses an exotic quantum phase
with a quadruply degenerate 
symmetry-protected topological ground state protected by a gap, 
in a quantum dot array with half-filled p-shell states combines the 
complexities of multiple carriers within one dot with that of multi-dot 
systems.

\begin{figure}
\includegraphics[width=0.99\linewidth]{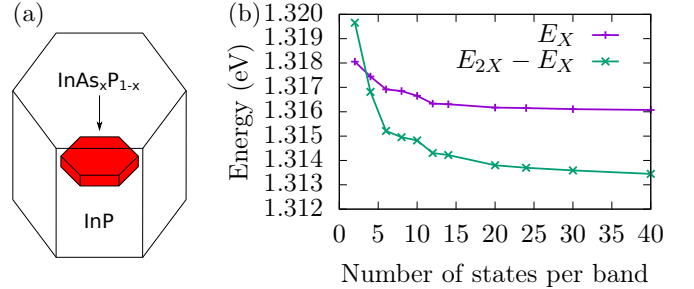}
\caption{\label{fig:dot_and_bx}(a) Hexagonal InAs$_x$P$_{1-x}$ 
quantum dot (red) inside a segment of an InP nanowire.
(b) Positions of the spectral lines emitted from the lowest biexciton and
exciton states as a function of the number of single-particle states in 
conduction and valence band used in a full CI calculation.}
\end{figure}

A quantitative theoretical description for such applications is highly desirable.
However, the direct numerical calculation of many-body states of multiple 
interacting carriers in a semiconductor nanostructure is difficult because of 
the curse of dimensionality, i.e. the fast growth of the many-body Hilbert space 
$\mathcal{H}$ with the number of carriers and single-particle states. 
For an excitonic complex composed of $n_e$ electrons and $n_h$ holes 
distributed on a set of $N_e$ confined electron and $N_h$ hole states, 
the dimension of $\mathcal{H}$ is 
${N_e \choose n_e}\times{N_h\choose n_h}$.
Dozens of single-particle states might be needed in full 
configuration interaction (CI) calculations\cite{CdSe_nanocrystals}, 
as shown in Fig.~\ref{fig:dot_and_bx}(b), which depicts
the positions of the spectral lines emitted from the lowest exciton 
and biexciton state as a function of the number of single-particle states 
$N_e=N_h$ for a hexagonal InAs$_{0.2}$P$_{0.8}$ quantum dot 
with diameter 18 nm and 
height 4 nm in a wurtzite InP nanowire\cite{QNANO_InAsP} as sketched in 
Fig.~\ref{fig:dot_and_bx}(a).
In particular, relative quantities like the biexciton binding energy 
$\Delta E_B=(E_{2X}-E_X)-E_X$ converge slowly because larger complexes 
generally converge more slowly than smaller complexes.

Therefore, for larger complexes full CI calculations become prohibitively 
demanding and one has to resort to approximate methods.
A common principle of many such approximations is that, in most situations,
only a small subspace of the full Hilbert space contributes significantly to 
the many-body states of interest, e.~g., to the ground state.
The Hartree-Fock method is a fast and easy approximation that seeks to find 
the optimal description in terms of a single configuration.
It may also help to speed up the convergence of larger CI calculations 
if they are performed starting from Hartree-Fock single-particle 
states\cite{HF_Abolfath}.
If the nanostructure possesses symmetries\cite{CI_FockDarwin},
the full Hilbert space can be decoupled into different blocks,
each of which has a much smaller dimension than the full problem.
Similarly, one may also exploit approximate hidden 
symmetries\cite{HiddenSymmetry, HiddenSymmetry_Exp} 
to investigate large excitonic 
complexes. 
In quantum chemistry, configuration interaction with single and double 
excitations (CISD) or coupled cluster (CC) 
approaches\cite{coulped-cluster_RevModPhys} are commonly used to calculate
many-body states in large Hilbert spaces.
Lately, also representations of many-body states in terms of
matrix product states (MPS) have been shown to perform 
well\cite{MPS_Cirac,MPS_Pfannkuche,CIPSI_MPS}, especially for
ground states and for one-dimensional systems.

In this article, we present a general method for the numerical 
calculation of correlated many-body states in quantum nanostructures that 
does not require any strong assumption about the wave functions such as
particular symmetries and also enables the calculation of excited states.  
In light of the analogy between quantum dots and atoms, it is suggestive to 
take inspiration from atomic and molecular physics, which specializes 
in interacting electrons.
Concretely, we implement a version of the 
CIPSI (configuration interaction by perturbation with
multiconfigurational zeroth-order wavefunction selected by iterative process)
method\cite{CIPSI_orig, CIPSI_Filippi2018, Heat-bath_Otten1,Heat-bath_Otten2} 
for the solution of problems involving interacting 
carriers in quantum nanostructures such as quantum dots, and demonstrate
its extreme efficiency.
This method consists of diagonalizing the many-body Hamiltonian in a reduced
space of configurations that are selected interatively by a criterion based 
on perturbation theory. After diagonalization in the relevant subspace, 
the effects of the configurations
outside of the selected subspace are accounted for perturbatively.
A major advantage of CIPSI is that it is a controlled approximation
as there exists a single convergence parameter, the threshold 
$\xi$, that controls the accuracy, where the
full CI result is obtained in the limit $\xi\to 0$. 
In quantum chemistry, selected CI methods have been applied to large, 
strongly correlated molecules, such as the chromium dimer, correlating 28 
electrons in 198 orbitals, leading to a total Hilbert space of 
$10^{42}$ using $10^{9}$ variational states and $10^{12}$ perturbative 
states\cite{Heat-bath_Otten3}.

Here, we describe the application of the CIPSI method
in the context of quantum nanostructures.
While typical applications in quantum chemistry aim at an accurate 
description of a few interacting atoms, 
modeling quantum nanostructures often involves 
hundreds of thousands to millions of atoms. In order to account for effects 
due to strain, alloying, and the underlying crystal structure, we perform 
tight-binding based atomistic calculations of single particle 
states\cite{QNANO_InAsP} as a starting point for many-body calculations.
We present numerical 
calculations to test the accuracy and numerical demands of the selected CI 
method for multi-exciton complexes. 
We find that the algorithm converges fast as 
a function of the threshold, 
so that results with near full CI accuracy are obtained
while an extremely small fraction of configurations of the full Hilbert space 
has been selected, reducing the numerical demands by many orders of magnitude.
A large part of the efficiency of the CIPSI algorithm can be attributed to the
perturbative corrections.

We then use the selected CI method to simulate the emission spectra
of three-exciton complexes, which requires the calculation of  
many excited biexciton states.
Subsequently, we investigate the role of the choice of the basis of the 
single-particle states and find that starting from the Hartree-Fock basis
can lead to a somewhat enhanced convergence for highly charged many-body 
complexes, but building Slater determinants from eigenstates of a
single-particle Hamiltonian turns out to be favourable for 
charge neutral systems. 

Finally, having found that the perturbative corrections are responsible
for a large part of the accuracy of the CIPSI algorithm and observing
that the perturbative terms only contain off-diagonal matrix elements,
we devise a method to accurately calculate many-body states that requires
the knowledge of only a small fraction of the Coulomb matrix elements
constructed from all single-particle states.
This is especially relevant when single-particle states are 
obtained from atomistic simulations, since the calculation of 
Coulomb matrix elements
is one of the most time-consuming steps in the overall 
procedure of the simulation of many-body states in quantum nanostructures.
Therefore, this approach, which potentially
reduces the total number of Coulomb matrix elements by orders of magnitude, 
is extremely useful by itself.

The article is structured as follows: First, we describe the theoretical 
background and the implementation of the selected CI method with
perturbative corrections. Then, we demonstrate the convergence for
ground states of complexes of up to 13 excitons. Subsequently, we 
apply the method to the simulation of three-exciton emission spectra, 
and, after discussing the role of single-particle basis states, 
we demonstrate how accurate calculations can be performed with a
limited set of Coulomb matrix elements.

\section{Theory}
\subsection{Tight-binding description of single-particle states}
The main goal of this article is to assess the efficiency and applicability of
a variant of the CIPSI algorithm for calculations of 
interacting carriers in quantum nanostructures.
As a specific example, we consider InAsP quantum dots in a wurtzite InP 
nanowire 
matrix as depicted in Fig. \ref{fig:dot_and_bx}(a). 
Such structures have been grown and investigated experimentally, e.g., 
in Ref.~\onlinecite{Dalacu_entangled_2014}.
In Ref.~\onlinecite{QNANO_InAsP},
we developed a method to simulate their electronic and optical
properties using our atomistic computational toolkit
QNANO~\cite{QNANO_InAsP,Zielinski_selfassembled_InAs,Marek_review,CdSe_rings}
suitable for large-scale parallelized million-atom 
calculations on a computer cluster.

At the core of the calculation of single-particle states 
is a description in terms of the tight-binding Hamiltonian
\begin{align}
H_{TB}=&\sum_{i=1}^{N_\textrm{at}}\sum_{\alpha=1}^{N_\textrm{orb}}
\epsilon_{i,\alpha} c^\dagger_{i,\alpha}c_{i,\alpha}
+\sum_{i=1}^{N_\textrm{at}}\sum_{\alpha,\beta=1} ^{N_\textrm{orb}}
\lambda_{i,\alpha,\beta} c^\dagger_{i,\alpha}c_{i,\beta}
\nn&
+\sum_{i=1}^{N_\textrm{at}} \sum_{j=1}^{nn(i)} \sum_{\alpha,\beta=1}^{N_\textrm{orb}}
t_{i,\alpha,j,\beta}c^\dagger_{i,\alpha} c_{j,\beta},
\label{eq:HTB}
\end{align}
where $c^\dagger_{i,\alpha}$ is the creation operator for an electron in the 
local orbital $\alpha$ on atom $i$. $N_\textrm{at}$ is the number of atoms 
in the sample and we use an spds$^*$-model with
$N_\textrm{orb}=20$ orbitals per atom. $\epsilon_{i,\alpha}$ are the onsite
energies, $t_{i,\alpha, j, \beta}$ are the nearest-neighbors 
hopping elements from orbital $\alpha$ on atom $i$ to orbital $\beta$ on 
atom $j$, and $\lambda_{i,\alpha,\beta}$ describes the spin-orbit coupling
at atom $i$.
To account for strain, the onsite and hopping parameters are modified
based on the local bond lengths and angles, which we obtain by
performing a valence-force-field strain relaxation. A detailed
description of the tight-binding parameters and strain corrections is 
given in Ref.~\onlinecite{QNANO_InAsP}.
The tight-binding Hamiltonian $H_{TB}$ is diagonalized, which yields the
energy eigenvalues as well as the single-particle eigenstates 
in terms of expansion coefficients of
a linear combination of the local orbitals.
%

\subsection{Many-body Hamiltonian}
Neglecting Auger processes\cite{CdSe_Auger}, 
which are strongly suppressed in gapped systems,
the many-body Hamiltonian for interacting electrons and holes in a
quantum nanostructure is
\begin{align}
H&=\sum_{i}E^{(e)}_i c^\dagger_i c_i +
\frac 12\sum_{ijkl} \langle ij|V_{ee}|kl\rangle c^\dagger_ic^\dagger_jc_kc_l
\nn&+ \sum_{p}E^{(h)}_p h^\dagger_p h_p +
\frac 12\sum_{pqrs}\langle pq|V_{hh}|rs\rangle h^\dagger_ph^\dagger_qh_rh_s
\nn& 
-\sum_{iqrl}\big(\langle iq|V_{eh}^\textrm{dir}|rl\rangle 
-\langle iq|V_{eh}^\textrm{exc}|lr\rangle\big) c^\dagger_ih^\dagger_qh_rc_l,
\label{eq:H}
\end{align}
where $E^{(e)}_i$ and $E^{(h)}_p$ are the single-particle energy eigenvalues
of the $i$-th conduction band electron state and 
of the $p$-th hole state (negative of the valence band electron energy 
eigenvalue), respectively, and
$c^\dagger_i$ and $h^\dagger_p$ are the corresponding creation operators
for electrons and holes.
$\langle ij|V_{ee}|kl\rangle$, $\langle pq|V_{hh}|rs\rangle$, 
$\langle iq|V_{eh}^\textrm{dir}|rl\rangle$, and 
$\langle iq|V_{eh}^\textrm{exc}|lr\rangle$ are the electron-electron,
hole-hole, as well as the direct and the exchange electron-hole 
Coulomb matrix elements, e.g.,
\begin{align}
\langle ij|V_{ee}|kl\rangle=\int d\mathbf{r}_1 \int d\mathbf{r}_2
\frac{e^2\psi^*_i(\mathbf{r}_1)\psi^*_j(\mathbf{r}_2)\psi_k(\mathbf{r}_2)
\psi_l(\mathbf{r}_1)}{{4\pi\epsilon \epsilon_0}|\mathbf{r}_1-\mathbf{r}_2|}.
\end{align}
For more details on the calculation of the Coulomb matrix elements
the reader is referred to Ref.~\onlinecite{QNANO_InAsP}.

A general many-body state in a semiconductor nano\-structure
composed of $n_e$ electrons and $n_h$ holes can be described by
\begin{align}
|\Psi\rangle=
\sideset{}{'}\sum_{\substack{\{\mu_1,\dots, \mu_{n_e}; 
\\ \nu_1,\dots, \nu_{n_h}\}}}
&A(\mu_1,\dots \mu_{n_e}; \nu_1,\dots, \nu_{n_h}) \times\nn&
c^\dagger_{\mu_1}\dots c^\dagger_{\mu_{n_e}}
h^\dagger_{\nu_1}\dots h^\dagger_{\nu_{n_h}} |0\rangle
\label{eq:conf}
\end{align}
where $\mu_i$ and $\nu_i$ denote indices of electron and hole states, 
respectively,
$|0\rangle$ is the semiconductor ground state with a full valence band 
and an empty conduction band, and $A$ are expansion coefficients.
The prime on the summation indicates that we sum only over indices
with $\mu_i < \mu_{i+1}$ and $\nu_i < \nu_{i+1}$.
A set of indices $\{\mu_1,\dots, \mu_{n_e}; \nu_1,\dots, \nu_{n_h}\}$
with the constraints $\mu_i < \mu_{i+1}$ and  $\nu_i < \nu_{i+1}$
defines a single configuration and all possible configurations together
form a complete basis of the many-body Hilbert space $\mathcal{H}$.
In order to keep the many-body Hilbert space finite, one typically only 
accounts for a finite number of $N_e$ electron and $N_h$ hole states. 

The full configuration interaction (CI) method consists of constructing 
all possible configurations 
$c^\dagger_{\mu_1}\dots c^\dagger_{\mu_{n_e}}
h^\dagger_{\nu_1}\dots h^\dagger_{\nu_{n_h}} |0\rangle$ in the expansion
of $|\Psi\rangle$ in Eq.~\eqref{eq:conf}
for a given number of electrons $n_e$ and holes 
$n_h$ and for a given number of single-particle states $N_e$ and $N_h$ 
and then solving the eigenvalue equation $H|\Psi\rangle=\lambda |\Psi\rangle$
to obtain the eigenvalues $\lambda$ and the eigenvectors in terms of
the expansion coefficients $A$.

\subsection{CIPSI Method}
In practice, the CI method is limited by the fact that the total 
dimension of the many-body Hilbert space $\mathcal{H}$ is given by
${N_e\choose n_e} \times {N_h\choose n_h}$,
so that a full CI treatment is only possible for a small number of interacting
carriers and single-particle states.
One method to tackle the analogous problem in the context of molecular physics
is the CIPSI method\cite{CIPSI_orig,CIPSI_Persico1987}.
There, the Hamiltonian is diagonalized only in a small subspace 
$\mathcal{H}_0\subset \mathcal{H}$ of the full many-body Hilbert space 
$\mathcal{H}$
corresponding to the most relevant states for the calculation.
Which states are selected as part of the relevant subspace is decided
iteratively by a criterion based on perturbation theory. 

The CIPSI method has the advantage that it is a controlled approximation as
there is a single convergence parameter, the threshold $\xi$, which defines
the accuracy. In the limit $\xi\to 0$
the full CI method is obtained, but the number of selected states
approaches Dim($\mathcal{H}$). For finite $\xi$,
only quantitatively important configurations are explicitly taken into account.
Additionally, the states that are not selected are taken into account 
by second-order perturbative corrections to the energy, which significantly
enhances the accuracy.

\begin{table}
\vspace{3mm}
\def\arraystretch{1.5}
\begin{tabularx}{\linewidth}{|lX|}
\hline
\multicolumn{2}{|l|}{ \textbf{CIPSI Algorithm:}} \\ \hline
(1) & Start with an initial set of selected configurations $\mathcal{H}_0$ \\
(2) & Diagonalize many-body Hamiltonian $H$ in $\mathcal{H}_0$. 
\\& Let $E^{(0)}_n$ be the $n$-th eigenvalue and $|n^{(0)}\rangle$ 
the corresponding eigenvector.\\
(3) & For given target eigenstates $|n^{(0)}\rangle$ and for 
configurations $|k^{(0)}\rangle$ outside of the selected state space
$\mathcal{H}_0$: \\&
Calculate 
$\xi_{nk}=\langle k^{(0)}|H| n^{(0)}\rangle/(E_n^{(0)}-E_k^{(0)})$ \\
(4) & If $|\xi_{nk}| > \xi$: Add the configuration $k$ to $\mathcal{H}_0$. \\
(5) & Repeat from step (2) until no new states are added in step (4). \\
(6) & Calculate the second-order perturbative corrections  \\&
$\Delta E^n_{PT}= \sum_{k \notin \mathcal{H}_0} 
|\langle k^{(0)}|H| n^{(0)}\rangle|^2/(E_n^{(0)}-E_k^{(0)})$.
\vspace{1mm} \\ \hline
\end{tabularx}
\caption{\label{tab:algorithm} Layout of the CIPSI algorithm.}
\end{table}

The algorithm is summarized in Tab. \ref{tab:algorithm}:
We start with an initial small subspace $\mathcal{H}_0$, possibly a single
configuration, and diagonalize the many-body Hamiltonian in the subspace 
$\mathcal{H}_0$. 
The resulting eigenstates, which we denote by $|n^{(0)}\rangle$, 
together with all configurations $|k^{(0)}\rangle$ outside
of $\mathcal{H}_0$ form a complete basis of the full many-body Hilbert space 
$\mathcal{H}$. 
In order to improve the accuracy of $|n^{(0)}\rangle$, we consider the 
interaction with the configurations $|k^{(0)}\rangle$ outside of 
$\mathcal{H}_0$ perturbatively. Recall that the first order 
perturbative correction $|n^{(1)}\rangle$ to the approximate eigenstate
$|n^{(0)}\rangle$ is
\begin{subequations}
\begin{align}
|n^{(1)}\rangle&= \sum_{k} \xi_{nk} |k^{(0)}\rangle,   \\
\xi_{nk}&=\frac{\langle k^{(0)}|H| n^{(0)}\rangle}{E_n^{(0)}-E_k^{(0)}},
\end{align}
\end{subequations}
where $E_n^{(0)}$ is the eigenvalue corresponding to the eigenstate
$|n^{(0)}\rangle$ of the Hamiltonian in the subspace $\mathcal{H}_0$
and $E_k^{(0)}=\langle k^{(0)} | H | k^{(0)}\rangle$ is the diagonal
energy of the con\-fi\-guration $|k^{(0)}\rangle$.

$\xi_{nk}$ can be understood as the contribution from 
the configuration $|k^{(0)}\rangle$ to the eigenstates $|n\rangle$ of
the full many-body Hamiltonian approximated by the state $|n^{(0)}\rangle$.
Therefore,
in the selected configuration interaction method 
CIPSI, a configuration $|k^{(0)}\rangle$ 
is considered to be important for a more accurate description of a 
target state $|n^{(0)}\rangle$ if $|\xi_{nk}|$ exceeds a given threshold
value $\xi$. Thus, we loop through the configurations $|k^{(0)}\rangle$
outside of $\mathcal{H}_0$ and, if $|\xi_{nk}|>\xi$, we add 
the configuration $|k^{(0)}\rangle$ to the selected state space 
$\mathcal{H}_0$ for the next iteration. 
This process is repeated until no new configurations
are selected, which typically requires five to ten iterations.

Note that, during the selection process, one already calculates 
all terms that enter the expression of the second order
perturbative correction to the energy eigenvalues
\begin{align}
&\Delta E^{n}_{PT}=\sum_{k}\frac{|\langle k^{(0)}|H| n^{(0)}\rangle|^2}
{E_n^{(0)}-E_k^{(0)}},
\end{align}
so that the perturbative energy corrections, which will be shown to 
improve the convergence significantly, can be obtained with no 
additional numerical effort.

For large system sizes the numerically most demanding part of the algorithm 
is the calculation of $\langle k^{(0)}|H|n^{(0)}\rangle$ in step (3).
This is due to the fact that the full many-body Hamiltonian $H$ 
connects states from $\mathcal{H}_0$ to a much larger space, 
henceforth denoted by
$\mathcal{H}_c$, which consists of all configurations obtained 
from configurations in $\mathcal{H}_0$ with additionally up to two 
excitations. In practice,
storing a vector of configurations in the large connected space 
$\mathcal{H}_c$ is the limiting factor of the algorithm in terms of 
memory consumption.

Furthermore, the represention of a vector in the connected space
$\mathcal{H}_c$ as a sparse vector in the full Hilbert space $\mathcal{H}$
requires searches in a list of size Dim$(\mathcal{H}_c)$.
Because the lookup is critical for the performance of the algorithm,
here, we implement it using hash tables, which have constant scaling
$\mathcal{O}(1)$ with respect to the length of the list, in contrast to,
e.~g., the search in an ordered list or in a binary tree that scales as 
$\mathcal{O}\big(\log \textrm{Dim}(\mathcal{H}_c)\big)$ or 
a brute-force search of a state in a list of states without pre-ordering 
which requires linear time in Dim$(\mathcal{H}_c)$.

The CIPSI algorithm described in Tab. \ref{tab:algorithm} is formulated
on the level of configurations, irrespective of the single-particle basis
from which the configurations are constructed. 
However, the choice of the basis states may influence the convergence
of the method.
Due to the strong confinement in quantum nanostructures like quantum dots, 
here, we choose to work most of the time in the basis of eigenstates of a 
single-particle Hamiltonian that captures the details of the structure, such
as the confinement potential, alloying, strain, 
and the underlying crystal lattice. 
In the present case, we use the spds$^*$ tight-binding 
Hamiltonian $H_{TB}$ described in Eq.~\eqref{eq:HTB}, 
but other effective single-particle methods 
like empirical pseudo-potentials\cite{Bester_InGaAs} might be
used as well.
Because in some scenarios CI calculations have been shown\cite{HF_Abolfath}
to converge faster using a single-particle basis consisting of
Hartree-Fock orbitals, we also test the convergence in the Hartree-Fock basis
in a later section.

Finally, we note that the CIPSI method allows us to treat ground and excited
states on the same footing. 
For example, for calculations of the lowest $n_{EV}$ states, we use the same
subspace $\mathcal{H}_0$ for all states. We diagonalize the Hamiltonian
in the subspace $\mathcal{H}_0$, take the $n_{EV}$ lowest eigenstates and
add in step (4) of the algorithm all configurations $|k^{(0)}\rangle$ 
to the selected state space for the next iteration if
$|\xi_{nk}|>\xi$ for any $n\leq n_{EV}$.

\section{Results}
We now test the CIPSI algorithm on the example of 
a hexagonal InAs$_{0.2}$P$_{0.8}$/InP nanowire quantum dot 
with a diameter of $18$ nm and a height of $4$ nm as depicted in
Fig. \ref{fig:dot_and_bx}(a). 
In the calculations we account for up to $N_e=N_h=40$ electron and hole states.

\subsection{Convergence of excitonic ground states}
\begin{figure}
\includegraphics[width=0.99\linewidth]{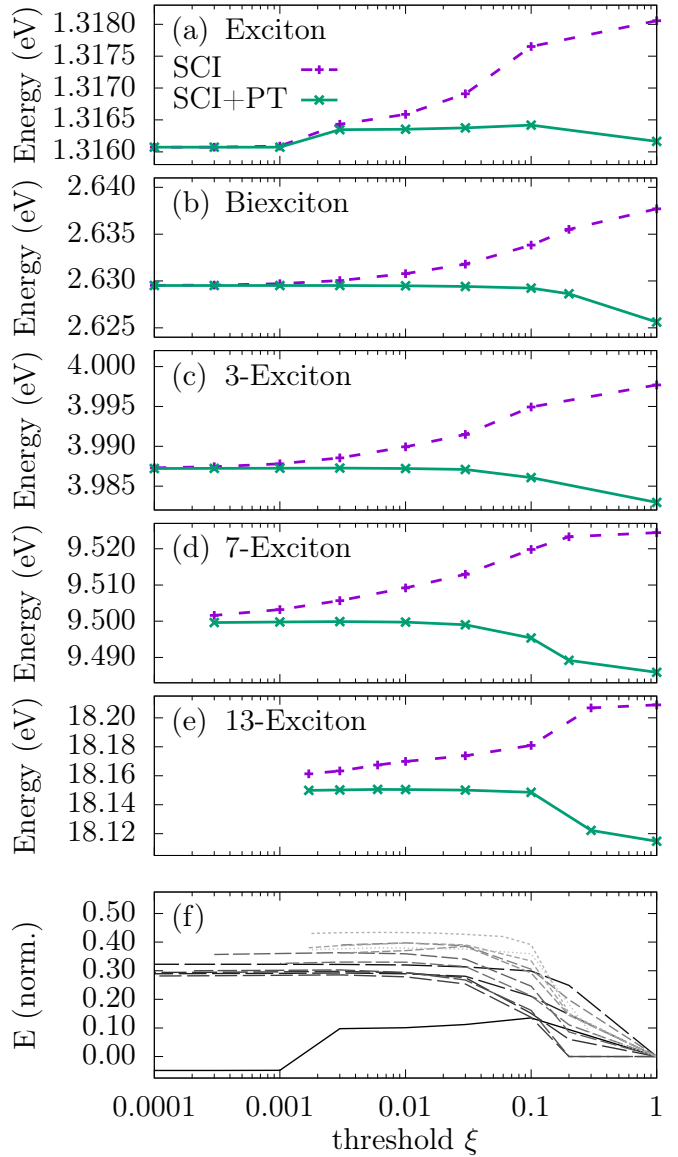}
\caption{\label{fig:convergence_multi} Energies of the ground states of 
complexes comprised of one (a), two (b), three (c), seven (d) and thirteen (e)
excitons as a function of the CIPSI threshold $\xi$. SCI (purple) denotes the
energy eigenvalue of the many-body Hamiltonian projected onto the subspace
of selected configurations, SCI+PT (green) includes the 
perturbative corrections from higher-energetic configurations.
(f) shows the ground states from one to thirteen excitons normalized according 
to Eq.~\eqref{eq:normalize}, where brighter lines with shorter dashes 
correspond to larger complexes.
}
\end{figure}

\begin{figure}
\includegraphics{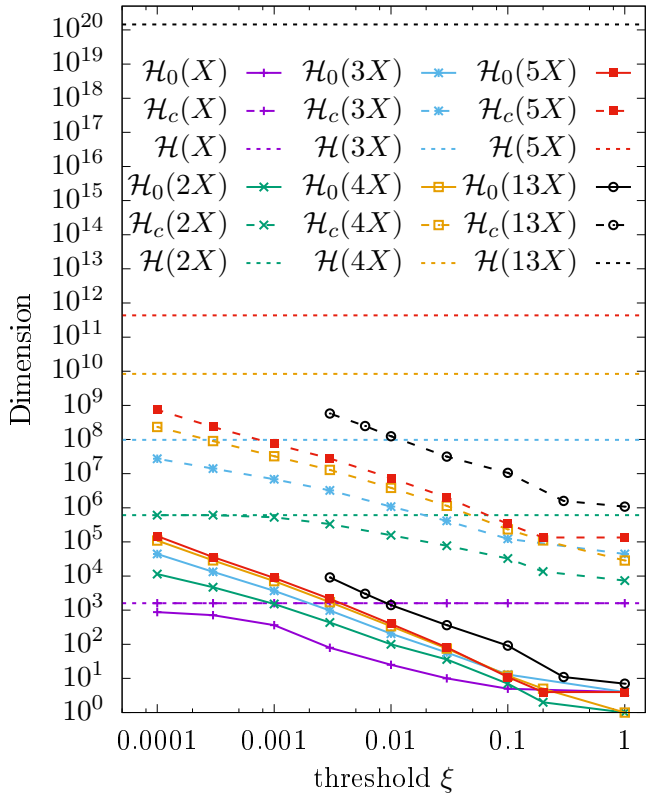}
\caption{\label{fig:convergence_dim} Number of selected configurations 
Dim($\mathcal{H}_0$), dimension of connected space Dim($\mathcal{H}_c$),
and dimension of the full many-body Hilbert space Dim($\mathcal{H}$) 
constructed from 40 electron and hole states as a function
of the threshold $\xi$ for the ground state calculation of excitonic
complexes composed of one to five and thirteen excitons.
}
\end{figure}

The lowest-energetic many-body states for systems consisting of one, two, 
three, seven and thirteen excitons calculated using the CIPSI method
with (SCI+PT) and without (SCI) second-order perturbative corrections are
depicted in Fig. \ref{fig:convergence_multi}(a-e) as a function 
of the threshold $\xi$. Note that smaller values of $\xi$ lead 
to the selection of more states and therefore correspond to results closer 
to full CI.
The convergence with respect to $\xi$ is qualitatively similar for most complexes,
although the energy scales are different. To highlight the general features,
we plot in Fig. \ref{fig:convergence_multi}(f) the CIPSI ground state energies
for all complexes up to 13 excitons normalized according to 
\begin{align}
&E_\textrm{normalized}^{nX}(\xi)=
\frac{ E_\textrm{SCI+PT}^{nX}(\xi) - E_\textrm{SCI+PT}^{nX}(\infty) }
{| \Delta E_\textrm{PT}^{nX}(\infty) |}.
\label{eq:normalize}
\end{align}
Here, $E_\textrm{SCI+PT}^{nX}(\xi)=E_\textrm{SCI}^{nX}(\xi)+
\Delta E_\textrm{PT}^{nX}(\xi)$ is the result of a CIPSI calculation 
of the gound state of the complex comprised of $n$ excitons for 
the threshold $\xi$ including the perturbative correction
$\Delta E_\textrm{PT}^{nX}(\xi)$. The reference energy scale 
$|\Delta E_\textrm{PT}^{nX}(\infty)|$ is given by the perturbative correction 
to the single configuration constructed from the lowest single-particle states,
where no configurations are added in step (4) of the algorithm.

Except for the single exciton, for which the full Hilbert space is  
comparatively small with Dim($\mathcal{H}$)=1600, 
the different excitonic complexes converge with respect to the
threshold $\xi$ in a similar way. The energy increases and reaches a
plateau at a threshold between $\xi=0.1$ and $\xi=0.01$. 
The value of the final energy indicates that the perturbative 
correction $|\Delta E_\textrm{PT}^{nX}(\infty)|$ from a single configuration
overestimates the influence of the remaining configurations by $30\%-50\%$.
The overestimation of the correction is a typical feature of perturbation 
theory in which higher order corrections often have alternating signs.

In order to assess the efficiency of the CIPSI algorithm for quantum
nanostructures, we plot in Fig.~\ref{fig:convergence_dim}
the number of selected states, i.e. the dimension of the subspace 
$\mathcal{H}_0$, the dimension of the connected subspace $\mathcal{H}_c$,
which determines the memory consumption of the algorithm,
and the dimension of the full many-body Hilbert space $\mathcal{H}$ 
constructed from up to 40 electron and hole states for calculations 
for excitonic complexes composed of one to five and of thirteen excitons.
We find that, except for the smallest complexes, the dimensions of the
different spaces differ by many orders of magnitude. 
For example, for the five-exciton complex at a threshold of 
$\xi=0.01$, one only has to diagonalize the many-body Hamiltonian in a space
with dimension Dim$(\mathcal{H}_0)\approx 400$ while perturbative corrections 
due to $7.4\times 10^6$ other configurations have to be performed 
out of the total many-body Hilbert space with the dimension
Dim($\mathcal{H}$) $\approx 4.3\times 10^{11}$.
It is noteworthy that, although the full many-body Hilbert space grows very fast
with the number of particles (about two orders of magnitude when one additional
exciton is added), the number of selected states as well as the dimension of 
the connected Hilbert space, which limits the calculations, 
increase much more slowly.
Therefore, the CIPSI algorithm is particularly useful for systems with a 
large number of particles.

\subsection{Convergence of optical spectra}
So far, we have only considered the convergence of the selected
CI algorithm for ground states, but it can equally well be
used to calculate excited states. This enables, e.g., 
calculations of emission spectra from higher excitonic complexes.
The spectroscopy of the three-exciton complex is particularly interesting,
as already the lowest three-exciton state requires the occupation of 
p-shell electron and hole states, 
whereas emission from single excitons and biexcitons after thermalization
predominantly originates from s-shell states. Therefore, from the spectral
lines emitted by the three-exciton complex one can infer information 
about the quantization and lateral confinement in quantum nanostructures. 
Although due to thermalization only a small number of three-exciton states
contribute to the spectrum, 
the simulation of transitions in the spectral range that 
contains three-exciton emission from the s-shell as well as from the p-shell 
is a good testing ground for selected CI calculations of excited states,
as it requires the calculation of a large number of excited biexciton states.

The optical emission spectrum of an excitonic complex can be described by 
Fermi's golden rule. 
The intensity of the emitted light with polarization direction
$\boldsymbol{\epsilon}$ is\cite{QNANO_InAsP}
\begin{align}
F(E,\boldsymbol{\epsilon})=F_0 \sum_{i,f} 
|\langle i| P(\boldsymbol{\epsilon}) |f\rangle |^2 
\delta\big[E-(E_f-E_i)\big] n_i(1-n_f),
\label{eq:fgr}
\end{align}
where $i$ and $f$ denote the initial and final many-body states, 
$E_i$ and $E_f$ are the respective energies, 
$\langle i| P(\boldsymbol{\epsilon}) |f\rangle$ is the 
dipole matrix element between states $i$ and $f$, $n_i$ and $n_f$ 
are the occupations of the initial and final states, and $F_0$ is
a constant depending on the light-matter interaction.

Here, we calculate the emission spectrum 
from three-exciton complexes to biexciton states, where we assume empty final
states $n_f=0$ and a thermal distribution of the initial three-exciton states
at a temperature $T=4$ K. 
The $\delta$-functions in Eq.~\eqref{eq:fgr} are broadened to Lorentzians
with a phenomenological linewidth of 0.1 meV. 
The many-body eigenstates
for both, initial three-exciton states and final biexciton states, are 
calculated using the selected CI algorithm and we add perturbative
corrections to the respective energy eigenvalues.
The dipole matrix elements are calculated from the many-body eigenstates 
following Ref.~\onlinecite{QNANO_InAsP}. 

\begin{figure}
\includegraphics{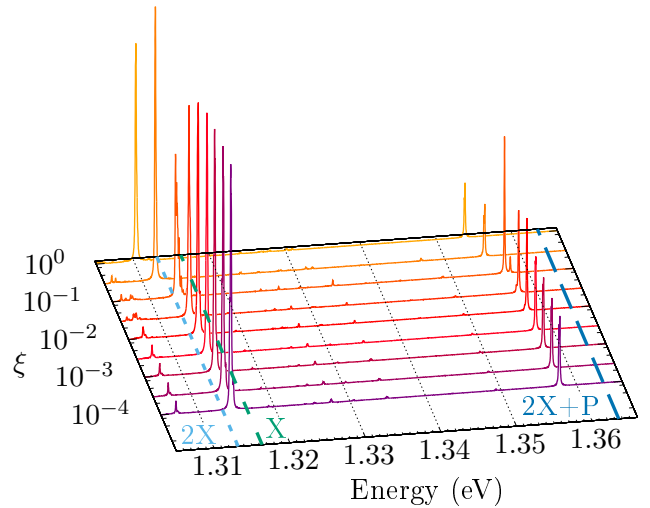}
\caption{\label{fig:spectra} 
Spectral lines emitted from the three-exciton complex for different
thresholds $\xi$. Additionally indicated are the positions of
the lowest bright exciton (X) and biexciton (2X) lines as well as 
the biexciton transtion energy shifted by electron and hole s-p-shell
splittings (2X+P).
}
\end{figure}

In Fig.~\ref{fig:spectra}, the
accumulated three-exciton emission spectrum for all polarization directions 
$F(E)=F(E,x)+F(E,y)+F(E,z)$ is depicted for different values of
the threshold $\xi$ in a spectral region that captures 
emission from s- and p-shell states.
As reference points we also mark the spectral positions of the 
bright emission lines for the s-shell exciton-to-ground-state (X) and 
biexciton-to-exciton (2X) transitions obtained from full CI calculations.
In addition, we indicate the energy obtained by adding the splittings between 
the lowest s- and p-shells for electrons and holes 
$\Delta E_p=[E^{(e)}_{1p}-E^{(e)}_{1s}]+[E^{(h)}_{1p}-E^{(h)}_{1s}]$
to the positon of the biexciton-to-exciton transition (2X+P).

Two peaks dominate the three-exciton spectra, one close to the lowest 
exicton transition, which corresponds to recombination of s-shell electrons
with s-shell holes, and one that is shifted by approximately the s-p-splitting
$\Delta E_p$, which stems from the recombination of p-shell electrons with
p-shell holes. Additionally, 
a number of very small peaks in the spectrum indicate dark states
that are optically forbidden either due to spatial symmetries, 
like recombination from p-shell electrons with s-shell holes, or spin 
selection rules.
Here, we find that the s-shell three-exciton transition line is found
between the biexciton and exciton lines. Furthermore, the distance between 
the two main peaks in the three-exciton spectrum 43.6 meV is about 
13\% smaller than the sum of the electron and hole s-p-splttings 
$\Delta E_p=$ 50.3 meV.
We attribute this significant deviation from the single-particle picture
to the fact that 
the spectral proximity of nearly degenerate p-orbitals makes it easier to 
reorganize charge densities to minimize Coulomb repulsion, 
so that the many-body contribution to biexcitons with p-shell 
carriers can be reduced compared to biexcitons with only s-shell carriers.
This finding implies that s-p splittings and confinement energies are
typically underestimated when they are derived
from the distance between three-exciton emission lines.

Regarding the convergence of the selected CI algorithm we find that, 
similar to the case of ground states of excitonic complexes discussed earlier,
the spectra are well converged at $\xi=0.01$.
Furthermore, note that, 
in order to fully capture the s-shell emission peak, more than 60 biexciton
states have to be calculated. To this end, calculations of the 120 lowest 
biexciton states have been performed. As we work with a single selected state 
space $\mathcal{H}_0$ for ground and excited states, a larger number of 
states are selected when 120 states are requested compared with the calculation
of only the ground state. For smaller values of the threshold $\xi$, however,
the ratio between the number of selected states in both cases is more and
more reduced and is found to be $\sim 10$ for $\xi=10^{-4}$. 

It is also noteworthy that we can make use of synergies in the calculation
of multiple eigenstates:
The approximate eigenstates $|n^{(0)}\rangle
=\sum_i \alpha_{ni}| i\rangle $ are stored as linear 
combinations of single configurations $|i\rangle \in \mathcal{H}_0$.
Then, the off-diagonal matrix element required in step~(3) of the
algorithm are calculated by
$\langle k^{(0)}|H|n^{(0)}\rangle=\sum_i\alpha_{ni}\langle k^{(0)}|H|i\rangle$.
For different eigenstates, only the coefficients $\alpha_{ni}$ change, but
the numerically costly matrix elements in terms of single configurations 
$\langle k^{(0)}|H|i\rangle$ have to be calculated only once. 
Therefore, calculating more eigenstates only leads to a marginal increase
in computation time, which makes the selected CI in practice very 
efficient for the calculation of a large number of eigenstates, as long as
enough memory for the simultaneous storage of $\xi_{nk}$ is available.

\subsection{CIPSI in Hartree-Fock basis}
\begin{figure}
\includegraphics{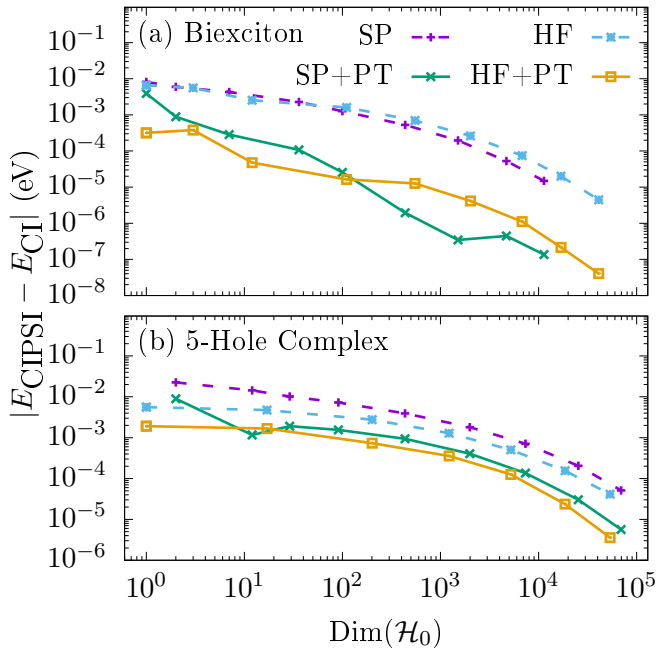}
\caption{\label{fig:HF}Convergence of CIPSI calculations
in the basis of single-particle eigenstates compared with calculations
starting from the Hartree-Fock basis for (a) the lowest biexciton and (b) the
lowest-energetic complex composed of five holes. The absolute error with 
respect to a full CI calculation is shown as a function of the number of
selected states on a double-logarithmic scale.
}
\end{figure}

In quantum dots charged with many electrons, it was shown\cite{HF_Abolfath} 
that CI calculations converge faster with the number of single-particle states
when configurations are constructed from Hartree-Fock single-particle states 
instead of eigenstates of a single-particle Hamiltonian. This is due to the 
fact that Hartree-Fock calculations already capture the redistribution of 
charge densities due to Coulomb repulsion.
To investigate whether also the convergence of the CIPSI method can 
be enhanced by working in the Hartree-Fock basis,
we present in Fig.~\ref{fig:HF} the absolute error 
$|E_\textrm{CIPSI} - E_\textrm{CI}|$ 
of the CIPSI algorithm with respect to the full CI calculation of the lowest
biexciton state [Fig.~\ref{fig:HF}(a)] and of the ground state of a 
many-body system comprised of 5 holes [Fig.~\ref{fig:HF}(b)] 
as a function of the number of selected states Dim$(\mathcal{H}_0)$ for 
calculations in the basis of eigenstates of the single-particle 
Hamiltonian (SP) as well as in the Hartree-Fock basis
(HF) with and without perturbative corrections (PT).

For the highly charged 5-hole complex, the calculation in the Hartree-Fock 
basis indeed generally leads to a smaller error for the same number of selected 
states. However, the errors in both bases are of the same order of magnitude,
in particular when more than a few states $\gtrsim 10$ are selected.
For the biexciton state, the first data point corresponds to 
a single configuration comprised of two electron and two holes in the lowest 
s-shells.
The energy of this state in the basis of eigenstates of the single-particle
Hamiltonian is $\sim 8.2$ meV above the full CI value taking into account 40
electron and holes states while a Hartree-Fock optimization of 
single-particle states reduces this value to $\sim 6.6$ meV.
It is noteworthy that including perturbative corrections 
to the single s-shell configuration in the basis of single-particle 
eigenstates already yields a more accurate result than the Hartree-Fock
calculation without corrections. With perturbative corrections, 
the respective Hartree-Fock state is found to be even more accurate by 
one order of magnitude.
Thus, the Hartree-Fock basis 
has a slight advantage over the single-particle Hamiltonian eigenstates 
when only a few states are selected.
However, reducing the threshold to select more states, we find that at
$\gtrsim 100$ states the single-particle eigenstates become
more favourable for convergence and the additional Hartree-Fock step 
required for the calculation is eventually detrimental.

\subsection{Reduced number of Coulomb matrix elements}
As we have shown so far, the CIPSI algorithm reduces the numerical demands 
for calculations of the many-body states of interacting carriers in quantum 
dots significantly compared to full CI calculations.
However, for practical purposes, a major problem that remains
is the calculation of Coulomb matrix elements. This is due to the fact that
the number of matrix elements, e.g., of $\langle ij|V_{ee}|kl\rangle$, scales
as $\mathcal{O}(N^4)$ with the number of single-particle states that are
accounted for in the calculation. 
Furthermore, even when only two-center terms are taken into account,
the calculation of a single Coulomb matrix element scales as 
$\mathcal{O}(N^2_\textrm{atoms})$ with the number of atoms $N_\textrm{atoms}$.
Therefore, the numerical demands of the calculation of Coulomb matrix element 
often limit the overall accuracy of the calculation.
One approach to attack this problem are linear scaling 
methods\cite{Zielinski_linear_scaling} with respect to $N_\textrm{atoms}$ to 
reduce the calculation time for a single matrix element.
Alternatively, one can speed up the calculation by 
vectorization\cite{Weidong_atomistic} or parallelization\cite{QNANO_InAsP}.

\begin{figure}
\includegraphics{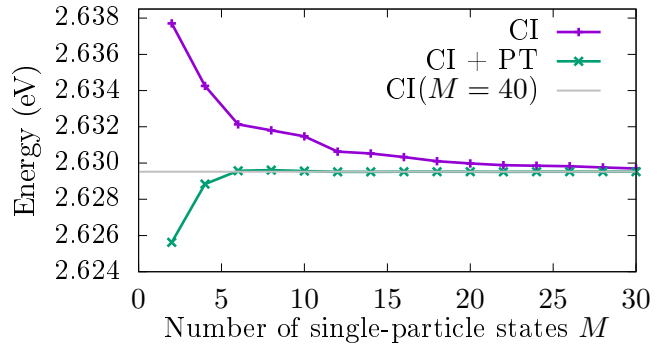}
\caption{\label{fig:restr}Full CI calculation of the lowest biexciton 
state using $M$ single-particle states with and without perturbative correction 
in the Hilbert space of up to 40 single-particle states.}
\end{figure}

Here, in light of the effectiveness of perturbative corrections, we propose 
another way to reduce the numerical demands for the calculation of 
Coulomb matrix elements:
Calculating many-body states in a restricted Hilbert space consisting
of at most $M$ single-particle states per band and adding the effects of 
a much larger Hilbert space with up to $N$ single-particle states 
perturbatively requires only the knowledge of matrix elements, e.g., 
$\langle ij|V_{ee}|kl\rangle$ with $i,j \le N$ and $k,l\le M$.
Thus, as long as the influence of configurations containing states $i>M$ is 
well described by perturbation theory, one only needs to calculate 
$\mathcal{O}(N^2M^2)$ instead of $\mathcal{O}(N^4)$ Coulomb matrix elements.

In Fig.~\ref{fig:restr} we show results of full CI calculations of the
lowest-energy biexciton states in a Hilbert space with up to $M$ 
single-particle states per band. Then, we add perturbative corrections
from higher-lying states with up to $N=40$ single-particle states per band
(CI+PT). It turns out that the calculation is practically converged at
$M=6$ states when perturbative corrections are included. 
Note that in this case only about $(M/N)^2\sim 2\%$ 
of all possible Coulomb matrix elements 
from $N=40$ single-particle states had to be used.
Therefore, using perturbative corrections not only speeds up
the calculation of many-body states, but also allows for a drastic reduction
of the computational resources for the Coulomb matrix element calculation 
without significantly reducing the overall accuracy.

\section{Conclusion}
We have presented a selected configuration interaction method with 
perturbative corrections that enables highly accurate and efficient 
calculations of many-body states of interacting charge carriers in 
million atom quantum nanostructures. 
This method has a number of advantages: It is controlled by a single 
convergence parameter $\xi$ and yields full CI results in the limit $\xi\to 0$, 
it is applicable in very general settings without requiring special 
conditions like symmetries, also allows an efficient calculation of 
excited states, and reduces the computational effort for obtaining
Coulomb matrix elements.
The fast convergence is demonstrated numerically for the ground states of 
complexes comprised of up to 13 excitons. We find that the calculations are
typically converged for thresholds $\xi$ between 0.01 and 0.1 
and within about 5 to 10 iterative state selection steps. 
A similar convergence is found of excited states, which we have tested 
by calculating the emission spectra from three-exicton complexes.
Finally, we have analyzed the choice of single particle states and
we have demonstrated a method for accurate many-body calculations
with a significantly reduced number of Coulomb matrix elements.

Our investigations show that, due to its extreme efficiency and accuracy, the 
selected configuration interaction method with perturbative corrections
can serve as a general purpose tool for the calculation of many-body states 
of interacting carriers in quantum nanostructures 
and it can yield quantitatively accurate results in
cases far out of reach for full configuration interaction calculations.
However, it is noteworthy that there exist optimized variants 
of the CIPSI method in the context of quantum chemistry\cite{CIPSI_Whaley2016}
that are even more efficient 
and it will be interesting to investigate and analyze their implementations for 
quantum nanostructure in the future.
In particular, the heat-bath CI variant\cite{Heat-bath_Otten1,Heat-bath_Otten2,
Heat-bath_Otten3,Heat-bath_cheap} offers great potential for accelerating
the state selection process by using a different selection criterion.
There, configurations $k$ are selected if 
$|\langle k| H | i\rangle c_i | < \epsilon$ for any $i$, where
$i$ and $k$ are single configurations, $c_i$ is the expansion coefficient 
of the target eigenstate in terms of the configuration $i$ and $\epsilon$
is an energy threshold. 
This criterion has the advantage that by pre-sorting the matrix elements 
a large number of non-contributing terms can be dropped in advance 
and do not have to be sampled explicitly. 
This paves the way for simulations of even larger systems of correlated 
electronic state in quantum nanostructures.

\acknowledgements
M.C. gratefully acknowledges funding from the Alexander-von-Humboldt
foundation through a Feodor-Lynen research fellowship.
P.H. acknowledges support from NSERC QC2DM Project and 
uOttawa Chair in Quantum Theory of Materials, 
Nanostructures and Devices.
This work was performed, in part, at the Center for Nanoscale Materials, a U.S. Department of Energy Office of Science User Facility, and supported by the U.S. Department of Energy, Office of Science, under Contract No. DE-AC02-06CH11357.
P.H. and M.C. acknowledge computational resources provided by Compute Canada
and by the Center for Nanoscale Materials.

\bibliography{qnano}
\end{document}